\documentclass[sigconf,natbib=true]{acmart}
\AtBeginDocument{%
  \providecommand\BibTeX{{%
    \normalfont B\kern-0.5em{\scshape i\kern-0.25em b}\kern-0.8em\TeX}}}

\copyrightyear{2023}
\acmYear{2023}
\setcopyright{acmlicensed}\acmConference[COMPASS '23]{ACM SIGCAS/SIGCHI Conference on Computing and Sustainable Societies}{August 16--19, 2023}{Cape Town, South Africa}
\acmBooktitle{ACM SIGCAS/SIGCHI Conference on Computing and Sustainable Societies (COMPASS '23), August 16--19, 2023, Cape Town, South Africa}
\acmPrice{15.00}
\acmDOI{10.1145/3588001.3609357}
\acmISBN{979-8-4007-0149-8/23/08}

\usepackage{natbib}  
\usepackage{caption} 
\usepackage{pgfplots}
\usepackage{siunitx}
\usepackage{amsmath}

\usepackage{multirow}
\usepackage{comment}
\usepackage{algpseudocode}
\usepackage{hyperref}
\usepackage{graphicx}
\usepackage{wrapfig}
\usepackage{algorithm}
\usepackage{pgfplots}
\usepackage{booktabs}
\usepackage{newfloat}
\usepackage{tcolorbox} 
\usepackage{listings}

\begin{document}

\title{Analysis of Elephant Movement in Sub-Saharan Africa: Ecological, Climatic, and Conservation Perspectives}


\author{Matthew Hines}
\email{matt_hines3@outlook.com}
\orcid{0009-0001-0710-0923}
\author{Gregory Glatzer}
\email{gregoryg323@gmail.com}
\orcid{0009-0006-5172-5652}
\author{Shreya Ghosh}
\email{shreya@psu.edu}
\orcid{0000-0002-6970-8889}
\affiliation{%
  \institution{College of IST, The Pennsylvania State University}
  \city{State College}
  \state{PA}
  \country{USA}
}

\author{Prasenjit Mitra}
\affiliation{%
  \institution{College of IST, The Pennsylvania State University}
  \city{State College}
  \state{PA}
  \country{USA}
}
\affiliation{%
  \institution{L3S Research Center, Leibniz University}
  \city{Hannover}
  \state{}
  \country{Germany}
}
\email{pmitra@psu.edu}
\orcid{0000-0002-7530-9497}

\renewcommand{\shortauthors}{Hines et al.}

\begin{abstract}
 The interaction between elephants and their environment has profound implications for both ecology and conservation strategies. This study presents an analytical approach to decipher the intricate patterns of elephant movement in Sub-Saharan Africa, concentrating on key ecological drivers such as seasonal variations and rainfall patterns. Despite the complexities surrounding these influential factors, our analysis provides a holistic view of elephant migratory behavior in the context of the dynamic African landscape.
Our comprehensive approach enables us to predict the potential impact of these ecological determinants on elephant migration, a critical step in establishing informed conservation strategies. This projection is particularly crucial given the impacts of global climate change on seasonal and rainfall patterns, which could substantially influence elephant movements in the future.
The findings of our work aim to not only advance the understanding of movement ecology but also foster a sustainable coexistence of humans and elephants in Sub-Saharan Africa. By predicting potential elephant routes, our work can inform strategies to minimize human-elephant conflict, effectively manage land use, and enhance anti-poaching efforts. This research underscores the importance of integrating movement ecology and climatic variables for effective wildlife management and conservation planning. Codebase available at \url{https://github.com/shreyaghosh-2016/Elephant-movement}
\end{abstract}

\ccsdesc[500]{Information systems~Information retrieval}
\ccsdesc[300]{Information systems~Data management systems}
\ccsdesc[300]{Computing methodologies~Machine learning}

\keywords{Movement ecology, trajectory analysis, spatio-temporal data}



\maketitle

\section{Introduction}

The intricate dynamics of elephant movement across Sub-Saharan Africa pose significant scientific interest and substantial conservation challenges~\cite{huang2022mapping}. These movements, driven by various ecological features such as water availability, temperature, and elevation, provide critical insights into their behavior, habitat preferences, and survival strategies~\cite{beirne2021african,branco2019determinants}. As such, a comprehensive analysis of elephant movements offers a valuable foundation for effective conservation planning and facilitates a more sustainable coexistence with these majestic creatures in their natural habitats.

Previous studies~\cite{evans2020natural,beirne2020climatic,de2019using} have outlined the role of water sources, temperature variations, and topography in influencing elephant movements. However, a comprehensive understanding necessitates the integration of additional factors such as seasonal and rainfall patterns, which have not been extensively examined to date. In the context of Sub-Saharan Africa, the two predominant seasons - the dry and wet seasons - manifest contrasting temperature, humidity, and rainfall conditions that are likely to significantly influence elephant migratory behavior. To address these gaps, our study employs a synergistic approach that combines \emph{Feature Engineering}, \emph{Clustering}, and \emph{Movement Modeling} techniques to analyze elephant movement data enhanced with elevation and water data obtained via Google's APIs. By utilizing the HDBSCAN clustering algorithm, our goal is to understand the relationship between the elephants' environmental variables and their behavior. Furthermore, we leverage the MoveHMM\footnote{\url{https://cran.r-project.org/web/packages/moveHMM/index.html}}, Trajr\footnote{\url{https://cran.r-project.org/web/packages/trajr/index.html}}, and MoveVis\footnote{\url{https://movevis.org/}} packages to conduct comprehensive modeling of elephant movement data.

One notable contribution of our work is the integration of climate change impacts into the analysis, recognizing their potential serious implications for future elephant travel routes. Understanding the shifts in seasonal patterns due to climate change is crucial in predicting likely changes in elephant movements and establishing effective conservation strategies. By incorporating climate change projections into our modeling, we can identify areas that may become more or less suitable for elephant populations, aiding in proactive conservation planning.

In addition to climate change, we consider rainfall patterns as a critical determinant of elephant movements. As elephants are known to adjust their travel routes according to water availability, a comprehensive analysis of rainfall patterns can provide predictive insights into future elephant migratory routes. This information is highly valuable, as it offers potential benefits to both human populations and elephants. Accurate predictions of elephant movements can guide local residents, ensuring minimal disruption to their lives, and inform conservation authorities to enhance protection measures against poaching threats.

Overall, this study proposes a multi-faceted approach to understanding elephant movement patterns, leveraging advanced computational techniques to decode the complex interplay of various ecological variables. By offering predictive insights into elephant movements, this research seeks to contribute significantly to the realms of movement ecology and sustainable conservation planning. By integrating climate change impacts and rainfall patterns into our analysis, we aim to provide a more comprehensive understanding of elephant movements and support informed decision-making for effective conservation strategies in Sub-Saharan Africa and beyond.
\section{Background}
\subsection{Related Works}
Understanding the complexities of elephant movement and migration has long been a subject of interest for ecologists and conservationists. Clements et al.~\cite{clements2014objective} developed a spatially explicit model of elephant movement in response to landscape heterogeneity and poaching risk, thereby providing a valuable foundation for the methodological approach used in our study. The authors demonstrated the importance of understanding landscape characteristics and threats to effective conservation strategy planning. Further insights into the effects of seasonal variations on elephant movement were provided by Loarie et al.~\cite{loarie2009elephant}. They emphasized the important influence of water availability on elephant migration, a factor closely linked to seasonal and rainfall patterns. They highlighted the urgency of understanding these migratory patterns in the context of climate change. Studies on human-elephant conflict, such as that by Fernando et al.~\cite{fernando2008review}, have underscored the need for informed conservation strategies that take into account potential elephant routes. This line of research reinforces the importance of our analytical approach, which predicts elephant routes and thus has direct applications for minimizing human-elephant conflict and enhancing anti-poaching efforts. The relationship between elephant movement and climate change, as investigated by Wall et al.~\cite{wall2014novel}, highlights the potential impact of changing seasonal and rainfall patterns on elephant migratory behavior. The role of climate change in altering the migration patterns of elephants was further studied by another work~\cite{kittle2016landscape}. They used GPS telemetry to track the movements of elephants in relation to changes in weather and climate conditions. Their research aligns with our study, furthering the understanding of how climate change impacts elephant migration. This understanding underscores the importance of our work in predicting these changes and establishing effective conservation strategies.
\par Naidoo et al.~\cite{naidoo2016estimating} proposed a model to analyze the impact of environmental features such as vegetation and water sources, as well as human threats, on the spatial distribution of elephants in Kruger National Park, South Africa. This work aligns with our study's aim to understand elephant movement in Sub-Saharan Africa in relation to ecological variables and threats. Boettiger et al.~\cite{boettiger2011inferring} developed an early warning system based on monitoring elephant movement and behavior in real-time to predict poaching threats. This work contributes to our understanding of using movement data to inform conservation strategies, a concept central to our study. The impact of climate change on elephant movement was investigated by Shannon et al.~\cite{shannon2009affects}. Their study emphasized the importance of understanding elephant response to rainfall variability, particularly in the context of climate change, a crucial component of our study. 
\par The impact of spatial factors on elephants' movement has been extensively explored in previous studies. Sitompul et al.~\cite{sitompul2013spatial} employed spatial data to determine that elephants tend to stay in proximity to the edge of canopy coverage. Mills et al.~\cite{mills2018forest} analyzed the influence of vegetation and seasonal variables on elephant movement. Human-Elephant Conflict (HEC) has also been a subject of investigation using spatial data. Tripathy et al.~\cite{tripathy2021descriptive} utilized KMeans clustering to identify hotspots of HEC. Sitati et al.~\cite{sitati2003predicting} examined the relationship between road proximity and crop raiding by elephants. Granados et al.~\cite{granados2012movement} focused on the influence of human infrastructure and settlements on elephant movement, particularly emphasizing seasonal patterns and crop raiding. In a broader context, Lamb et al.~\cite{lamb2020space} explored the application of clustering algorithms to elucidate groupings within animal movement data.

While our work does not directly address HEC, it contributes to the existing body of knowledge by employing multiple clustering algorithms to identify human settlements that elephants seem to inhabit. Through spatial analysis, we aim to provide insights into the spatial distribution of elephant presence, thus adding to the understanding of their movement patterns and potential interactions with human settlements. Animal movement studies, especially of large mammals like elephants, are of considerable interest to ecologists, conservationists, and data scientists. Current research on elephant movement primarily considers key ecological factors like water availability, temperature, and elevation to understand elephant behavior and predict their migratory routes. The interest for deeper understanding has led to the utilization of advanced analytical techniques, including feature engineering and clustering algorithms in data science, to enrich elephant data and discern patterns. Simultaneously, movement modeling tools like MoveHMM, Trajr, and MoveVis have been used to build robust models of elephant movements, providing insights into their behavior based on environmental factors. However, these studies largely focus on a narrow set of ecological variables, with the influence of broader climatic phenomena such as seasonal and rainfall patterns, and global warming impacts on elephant movements relatively unexplored. Our study expands upon this existing knowledge, incorporating seasonal variations, rainfall patterns, and global warming effects to enhance predictability of elephant movements, contributing to a more sustainable coexistence of humans and elephants in Sub-Saharan Africa.
\vspace{-0.3cm}
\subsection{Problem Formulation}
The overarching problem addressed by our research pertains to the comprehensive understanding and prediction of elephant movement patterns across Sub-Saharan Africa. This problem can be broken down into the following specific challenges:

\textbf{Seasonal Influence on Elephant Movements:} While previous studies have analyzed various determinants of elephant movement, including water availability, temperature, and elevation, the influence of seasonal changes remains largely unexplored. We hypothesize that the contrasting weather, humidity, and rainfall conditions in the dry and wet seasons significantly impact elephant migration behavior. However, the relationship between seasonal patterns and elephant movements needs rigorous empirical investigation.

\textbf{Impacts of Rainfall Patterns:} Elephants are known to migrate in response to water availability, which is significantly influenced by rainfall. Yet, the role of rainfall patterns, including their temporal and spatial distribution, in determining elephant movement routes remains unclear. A comprehensive analysis of these patterns can potentially enhance the predictability of elephant movements.

\textbf{Effects of Climate Change:} Global warming is likely to cause shifts in seasonal patterns and rainfall distributions, which could significantly alter elephant migration routes. Understanding these impacts requires the integration of climate change projections with the analysis of elephant movements. The challenge lies in the complexity and uncertainty associated with climate change impacts and their translation into elephant movement behaviors.

\textbf{Predicting Elephant Movements and Analysing Movement Charecteristics:} Given the various factors influencing elephant movements, predicting their future routes is a highly complex problem. Accurate predictions are essential for informing conservation strategies and facilitating human-elephant coexistence.
We propose a data-driven approach, combining feature engineering, clustering, and movement modeling techniques that can help to predict elephant movements based on an expanded set of ecological variables. The resolution of these challenges would significantly enhance our understanding of elephant behaviors and contribute to more effective conservation planning.
\section{Proposed Framework}

This work follows a three-step approach to understanding elephant movements in Africa, employing Feature Engineering, Clustering, and Movement Modeling.
\subsection{Feature Engineering}
This initial step involves the integration of elevation and water body data into the existing elephant data. The elevation data is obtained via the Open Elevation API and is combined with each instance of the elephant data. On the other hand, the water body information is sourced in the form of polygon shapefiles. These shapefiles are processed using the nearbyWaterBodies() function, which generates edge points of each polygon shapefile. This information is subsequently utilized by the KMeans algorithm to identify prominent water bodies. 
Here, the elevation data is acquired and combined with each data point of the existing elephant data. Let's denote the data point as P and the elevation data as E. Each data point P(i) is combined with its corresponding elevation data E(i).
\begin{equation}
    P(i)' = P(i) 	\cup E(i)
\end{equation}
Where P(i)' is the new data point.
\subsection{Clustering}
Spatial movement data derived from animals in real-world scenarios often exhibit complex patterns. For instance, elephants have been observed to engage in shuttling motion, fluctuating between shaded areas and water bodies. Additionally, they can traverse significant distances over time without intermittent pauses or resting spots. Such nature of movement necessitates a strategy to distinguish between the shuttling motion and "locations of interest" within the spatial data. These locations of interest can be determined as regions where multiple data points congregate into a dense cluster, indicating frequent returns or prolonged stay by the elephants. Conversely, sporadic, thinly scattered points, symbolizing elephants' motion towards their next rest area, should be recognized as noise.

To identify these locations of interest, we resort to clustering. Considering the shuttling behaviour of elephants, a critical requirement is the ability to differentiate clusters from noise. The DBSCAN algorithm lends itself as a suitable candidate for this task owing to its inherent concept of noise, a feature absent in many popular clustering algorithms such as K-Means. We utilized the scikit-learn implementation of DBSCAN, incorporating a preprocessing step using scikit-learn's StandardScaler.
The refined data set (after pre-processing) is subjected to clustering analysis, utilizing Hierarchical Density-Based Spatial Clustering of Applications with Noise (HDBSCAN) and Agglomerative Clustering (AGGLO) algorithms.

\emph{(1) HDBSCAN:} An enhancement of the DBSCAN algorithm, HDBSCAN allows for the accommodation of clusters with varying densities. This algorithm replaces the conventional epsilon parameter with min\_cluster\_size, enabling easier optimization by defining the smallest cluster size of interest in the study.

\emph{(2) AGGLO:} AGGLO is particularly effective in handling elongated and thin clusters. However, unlike HDBSCAN, AGGLO incorporates all noise present in the data set.

In the HDBSCAN algorithm, it creates clusters by connecting areas of the dataset where the distance between points is lower than a threshold. In the case of AGGLO, it's a type of hierarchical clustering where it starts with each point as a separate cluster and merges clusters based on a distance metric.

Let's denote $d(x, y)$ as the distance between points $x$ and $y$. For HDBSCAN, it looks for $x, y$ such that $d(x, y) < \epsilon$, where $\epsilon$ is a threshold distance. For AGGLO, it starts with each point as its own cluster and repeatedly merges the closest pair of clusters.
\subsection{Temperature Influence in Clustering}
In light of the findings from existing research works, underscoring the significance of temperature in explaining the shuttling motion of elephants, we incorporated temperature into our analysis. Accordingly, we executed DBSCAN under two feature spaces: (1) Temperature-influenced - comprising temperature, latitude, and longitude, and (2) Without Temperature influence – encompassing only latitude and longitude. The latter feature space excludes the temperature feature to evaluate its impact on the clustering algorithm's efficiency in identifying locations of interest.
\subsubsection{Obtaining Historical Temperature Data}
However, such temperature analysis is only feasible when temperature data is available. Certain studies concerning elephant movement (Tsalyuk et al.~\cite{tsalyuk2019temporal}; Wall et al.~\cite{wall2014data}) lack a temperature feature. Therefore, we explored methods to approximate temperature data from other data sources. Using the meteostat python package and API, we identified weather stations proximate to the study site. The historical data was queried and appended to the study data, enabling calculation of Temperature-influenced centroids that would have been impossible to calculate otherwise.

The procedure entailed three key steps: (1) Identifying a nearby weather station, (2) Matching timestamps with the queried data, and (3) Evaluating the capability of the appended historical temperature data in calculating temperature-influenced centroids.

For the first step, the median latitude and longitude of the given elephant's movement data was computed, which was then used to query a nearby station. The second step involved normalizing and interpolating the time series data from the station, provided by meteostat, to ensure a higher temporal granularity that matches the given data. In the third step, the correlation between the historical station data and Kruger temperature data was evaluated using the coefficient of determination, R-squared.

Our results indicate a moderate correlation between the study data and the station data. This correlation, combined with the performance of the Temperature-influenced centroids with the weather data, gives us confidence to extend this technique to datasets that lack temperature data.
Based on our experiment with elephant AM306 (See Figure~\ref{tempclust1}) from the Kruger dataset, we found that the Temperature-influenced feature space aided in revealing more nuanced locations of interest within the larger clusters identified by the Without Temperature influence feature space.
\subsubsection{Fuzzy Timestamp Matching}
Fuzzy timestamp matching is an advanced data processing technique that matches timestamps not based on exact equality but within a certain tolerance level. This tolerance level, or fuzzy threshold, is usually calculated by taking half of the median of the difference of timestamps in the dataset. The mathematical representation of the fuzzy timestamp matching process could be described as follows:\\
Given two timestamps, $t1$ and $t2$, and a tolerance level $\delta$, the timestamps $t1$ and $t2$ are said to match if:
\begin{equation}
    |t1-t2| \leq \delta
\end{equation}
where $|t1 - t2|$ denotes the absolute difference between the timestamps $t1$ and $t2$. In this case, $\delta$ is calculated as:
\begin{equation}
    \delta = 0.5 * median(|t[i+1] - t[i]|), \forall i \; 1 \; to \; N-1
\end{equation}
where N is the total number of timestamps, and t[i] represents the ith timestamp in the ordered sequence. This fuzzy matching approach increases the likelihood of matches and can help to mitigate data loss when aligning data from different sources or with different temporal resolutions. However, it is important to note that this technique may also introduce some uncertainty into the analysis due to the mismatched timestamps. Hence, an appropriate balance between data retention and accuracy should be maintained while deciding the value of $\delta$.
\par The integration of weather station temperature data with animal movement datasets presented a significant challenge due to the relatively low percentage of matching timestamps. For instance, in the case of AM189 from Etosha, a mere 19.662\% of timestamps corresponded. This limited overlap signifies a considerable loss of data, which undermines the analysis. To address this issue, we utilized "fuzzy" timestamp matching. This method extends the criteria of a match beyond exact timestamp equality, incorporating a pre-defined threshold for the discrepancy between two timestamps that still qualifies them as a match. The mathematical formulation of this concept is as follows:
\begin{figure}
    \centering
    \includegraphics[width=0.5\textwidth]{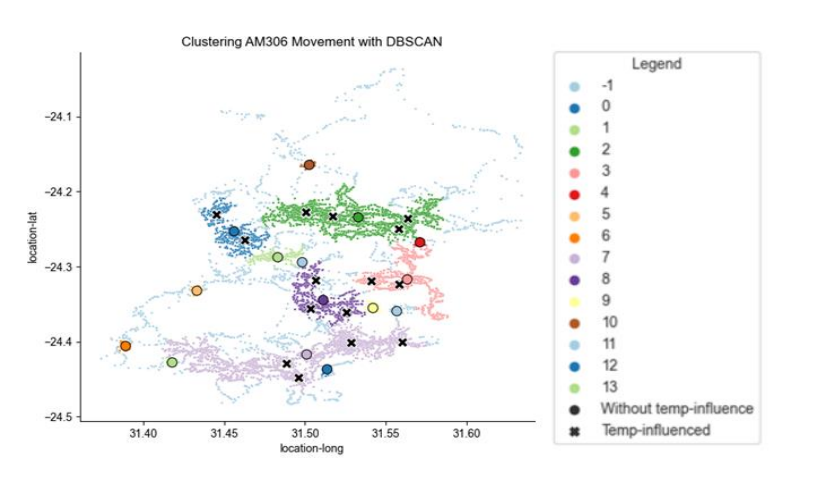}
    \caption{Comparing feature spaces with and without temperature on elephant AM306 from the Kruger dataset. Black Xs represent centroids calculated with temperature included in the feature space. Large colored dots represent centroids of clusters calculated solely on location data. Parameters: Without temp-influenced epsilon=0.1, minPts=35. Temp-influenced epsilon=0.2, minPts=50}
    \label{tempclust1}
\end{figure}
\begin{figure*}
    \centering
    \includegraphics[width=0.9\textwidth]{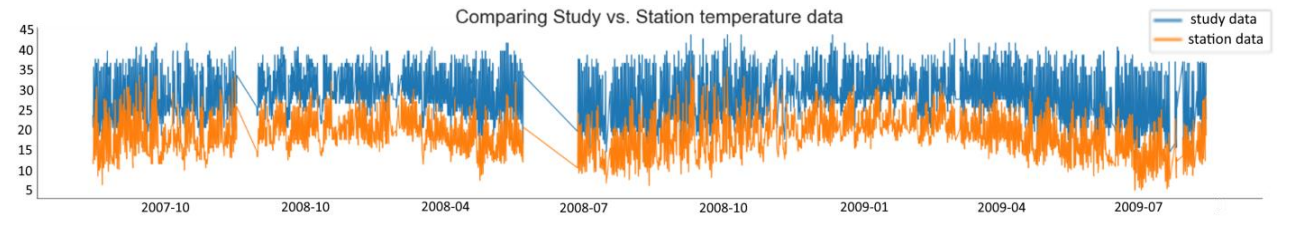}
    \caption{The difference between temperature data from Elephant AM105 in the Kruger dataset, and
corresponding temperature data from the meteostat weather station API.}
    \label{fuzzyres}
\end{figure*}
\begin{table}[htbp]
\centering
\renewcommand{\arraystretch}{1.2}
\begin{tabular}{@{}lr@{}}
\toprule
\textbf{Metric} & \textbf{Value} \\
\midrule
R-squared (zero-centered) & 0.6871044690549571 \\
Offset (study -- station) & 9.840106696689293 \\
\% of timestamps found & 61.6\% \\
\bottomrule
\end{tabular}
\caption{Statistics for Figure~\ref{fuzzyres}}
\label{tab:metrics}
\end{table}
Given two timestamps t1 and t2, and a tolerance level (or fuzzy threshold) $\delta$, the timestamps t1 and t2 are said to match if the absolute difference between them, denoted as |t1 - t2|, does not exceed $\delta$. The fuzzy threshold $\delta$ is calculated as half the median of the differences between all sequential pairs of timestamps in the dataset.

By employing fuzzy timestamp matching, the percentage of matched data can be substantially increased. For example, in the case of AG191 from Etosha, conventional timestamp matching resulted in a match percentage of 41.85\%. With the application of fuzzy matching, this percentage rose to 74.50\%.
\begin{figure}
    \centering
    \includegraphics[width=0.5\textwidth]{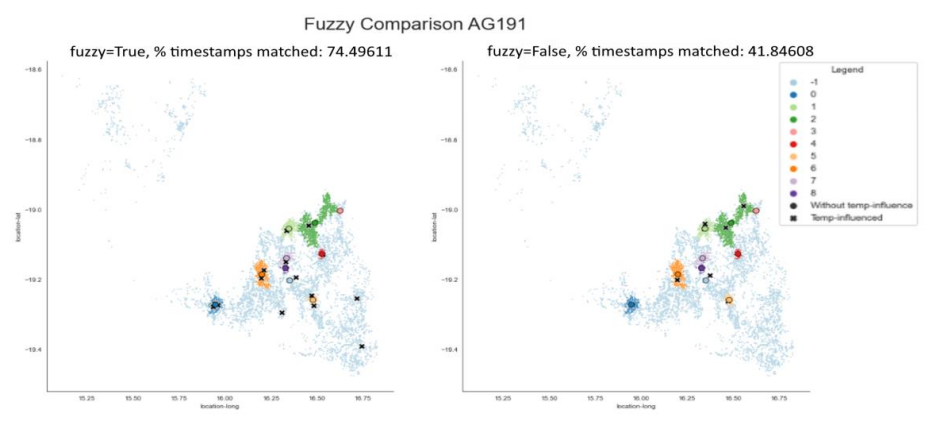}
    \caption{Fuzzy timestamp matching results on AG191 from the Etosha dataset
}
    \label{fuzzy}
\end{figure}
Notably, these additional matches, achieved through fuzzy matching, are proximal rather than exact timestamp correlations, falling within the defined tolerance. The increased density of matched timestamps facilitated the identification of the same temperature-influenced centroids as found using precise matching, along with additional centroids, particularly in the lower-right region of the map (See Figure~\ref{fuzzy}).

Subsequent trials of fuzzy timestamp matching yielded consistent results, enhancing the number of temperature-influenced centroids identified, particularly in regions that initially lacked such centroids. Considering these improvements, it is recommended that fuzzy timestamp matching be implemented in future endeavors to identify centroids in animal movement data.

Figure~\ref{fuzzyres} and Table~\ref{tab:metrics} present the outcomes of conducting steps (1)-(3) on Elephant AM105 from the Kruger dataset. When comparing the weather station temperature data with the study temperature data, we find a moderate correlation, as indicated by an R-squared value of 0.6978. As stated by Moore et al.~\cite{moore2009introduction}, an R-squared value between 0.5 and 0.7 typically signifies a moderate effect size. Prior to calculating the R-squared value, the data was zero-centered, which normalizes the data around zero. Nevertheless, this process is not critical for our analysis because we're primarily interested in the correlation between temperature values for clustering, rather than the individual temperature values themselves. It's vital to pay attention to the "\% of timestamps found" data. In this instance, temperature measurements from the weather station coincide with only 61\% of the original dataset for Elephant AM105. This level of data retention, achieved after applying both normalization and interpolation techniques, suggests significant data loss.
\begin{figure}
    \centering
    \includegraphics[width=0.5\textwidth]{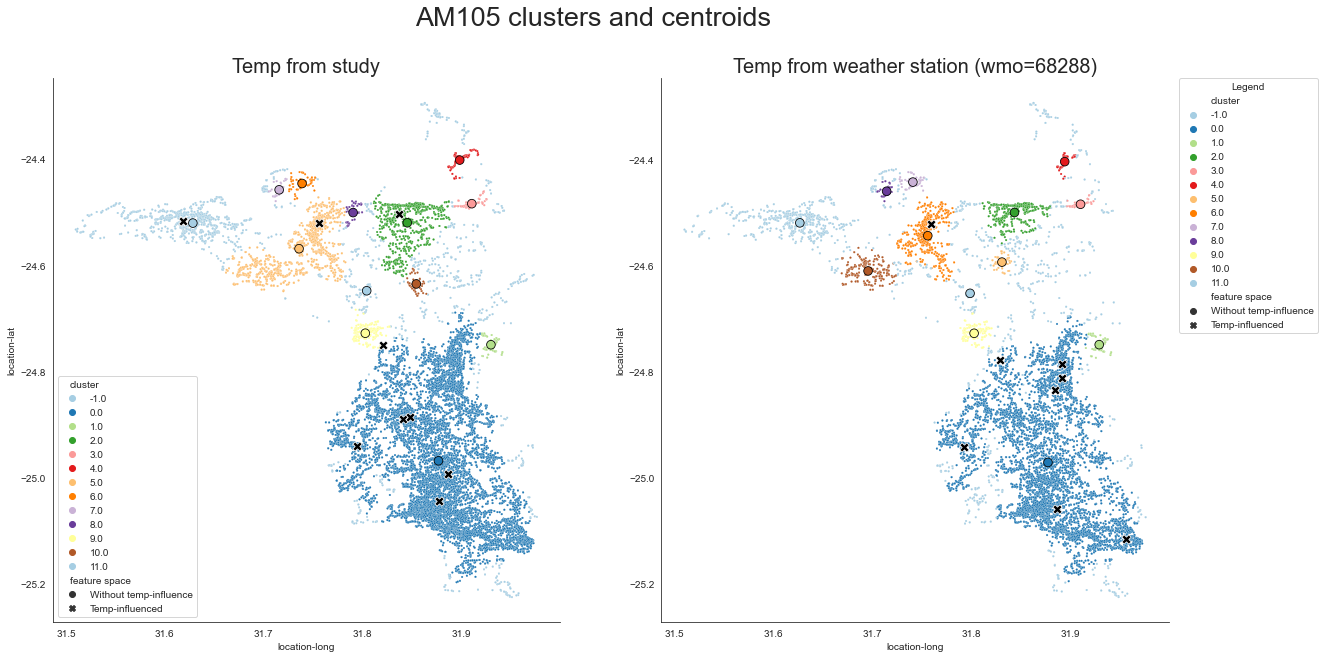}
    \caption{Applying DBSCAN to AM105 with study data and weather station data}
    \label{KrugerWeatherStationVsStudy}
\end{figure}
One potential solution to this issue is fuzzy timestamp matching, which aims to match timestamps within a certain threshold. While this method can increase data retention, it may also compromise data quality in favor of quantity. The application of fuzzy timestamp matching will be discussed further in the subsequent sections. Finally, we examine the performance of DBSCAN clustering when using weather station data. After examining the performance of DBSCAN on the elephant AM105 weather data in the Kruger dataset (See Figure~\ref{KrugerWeatherStationVsStudy}), it is evident that the Temp-influenced centroids effectively detect areas of high density, indicating the previously mentioned points of interest. Remarkably, the station data not only identifies some of the same points of interest as the original study data but also reveals new locations. It is important to note that identical parameters were used for both analyses. Given the successful performance of the Temp-Influenced centroids technique on the weather data and the moderate correlation observed between the study data and the station data, we can now proceed with greater confidence to apply this methodology to datasets lacking temperature information. This opens up new possibilities for exploring and analyzing diverse datasets using this technique.
\subsection{Movement Modeling}
Finally, movement modeling is conducted on the processed elephant data using the MoveHMM, Trajr, and MoveVis packages in R.

\emph{(1) MoveHMM Model:} MoveHMM package employs a Hidden Markov Model with a bivariate time series to model an animal's step length and turning angle at each location. For our study, two states are defined for the elephants. The first state, foraging, is characterized by small step sizes and larger turning angles. The second state, transiting, exhibits large step sizes and small turning angles.
\begin{figure}
    \centering
    \includegraphics[width=0.5\textwidth]{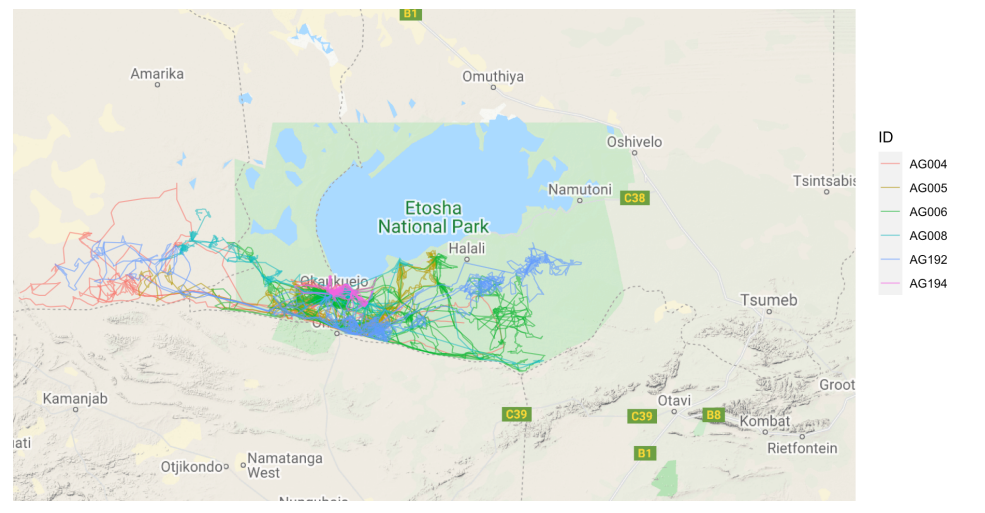}
    \caption{Etosha Mapping For All Elephant Groups}
    \label{fig:5}
\end{figure}

\emph{(2) Trajr Model:} The Trajr package uses first and second order finite derivatives to calculate velocity and acceleration from a Trajectory Object. Additional features such as straightness and directional change can also be computed, offering insight into sudden speed changes and areas of limited directional change.
\begin{figure}
    \centering
    \includegraphics[width=0.5\textwidth]{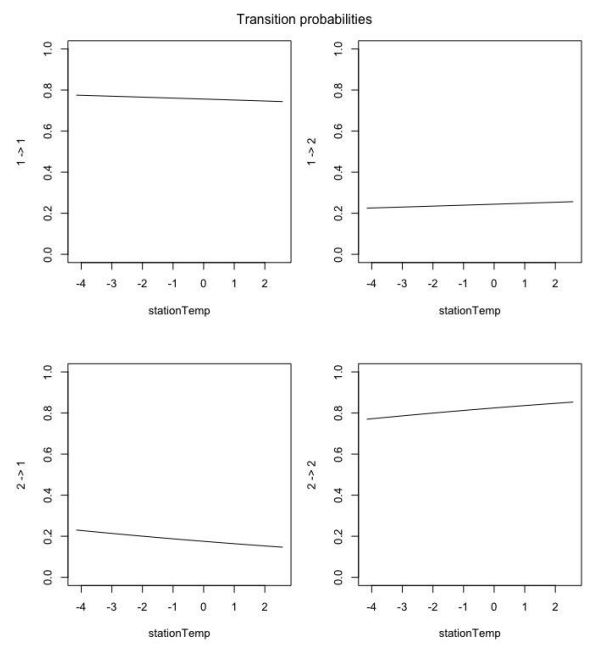}
    \caption{Etosha Transition Probabilities}
    \label{fig:6}
\end{figure}
\begin{figure}
    \centering
    \includegraphics[width=0.5\textwidth]{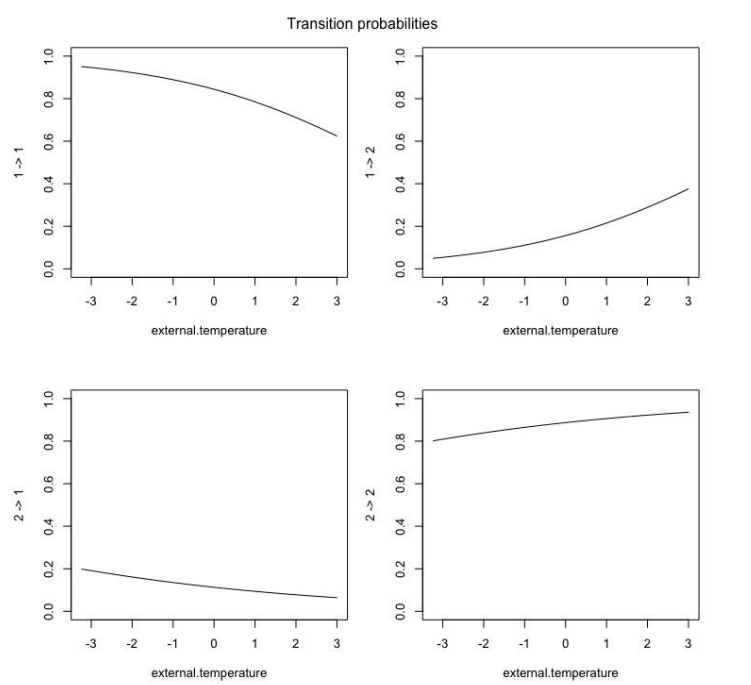}
    \caption{Kruger Transition Probabilities}
    \label{fig:7}
\end{figure}

\emph{(3) MoveVis Model:} The MoveVis package facilitates the animation of movement data. Spatial frames are created from a move object, which is generated using a built-in function that reads the elephant data. The resultant animated visuals provide an intuitive understanding of elephant movement patterns.
\begin{figure*}[htbp]
    \centering
    \includegraphics[width=0.7\textwidth]{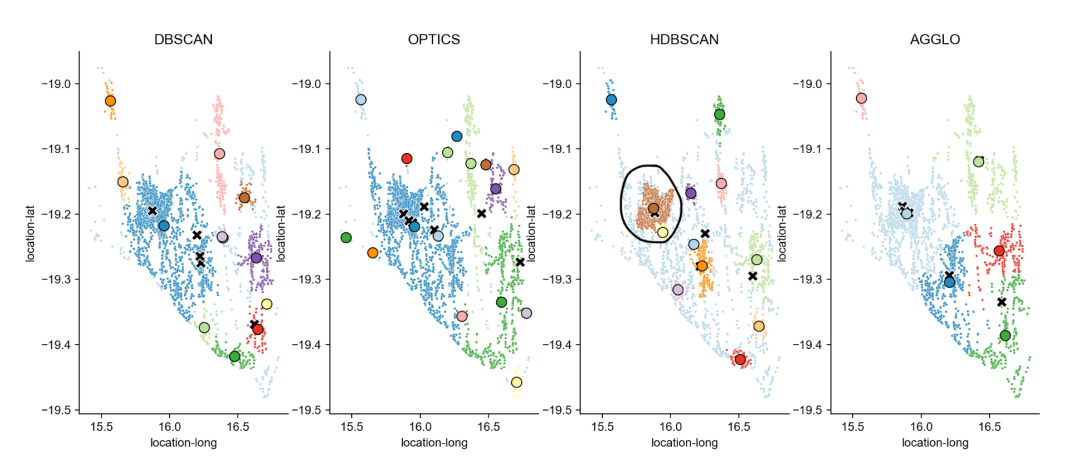}
    \caption{Clustering with Temperature on HDBSCAN and AGGLO AG006}
    \label{fig:1}
\end{figure*}
In the MoveHMM model, it uses a Hidden Markov Model. The model assumes that there is an underlying, unobservable state process that governs the dynamics of the observable process. The state at time $t$, $S(t)$, is only dependent on the state at time $t-1, S(t-1)$, which is represented as $P(S(t)|S(t-1))$.

In the Trajr Model, it calculates velocity and acceleration of an animal by the Trajectory Object. Let $P(i)$ be the position at time $i$, $V(i)$ be the velocity, and $A(i)$ be the acceleration. These can be calculated as:
\begin{equation}
\begin{split}
       V(i) = P(i) - P(i-1) \\
A(i) = V(i) - V(i-1)
\end{split}
 \end{equation}
Where the subtraction of position values gives the velocity, and the subtraction of velocity values gives the acceleration.

In the MoveVis model, it creates spatial frames from a move object, where each spatial frame F(t) at time t is a snapshot of all the position data at that time, which can be denoted as:
\begin{equation}
   F(t) = \{P(i) : i \in T(t)\} 
\end{equation}
Where $T(t)$ are all the indices at time $t$.

Each of these methods in the pipeline serves a distinct yet complementary role in elucidating the factors influencing elephant movement patterns across Africa, thereby contributing to more effective conservation strategies.

\section{Experimental evaluation}
\subsection{Performance Metrics for Clustering}
The effectiveness of the two clustering algorithms, HDBSCAN and AGGLO, was assessed using the elephant data that included temperature and elevation. Figure~\ref{fig:1} illustrates the clustering results obtained from these methods for the elephant ID ``AG006". In comparison to the DBSCAN algorithm, HDBSCAN demonstrated the ability to form varying density clusters. 
\begin{figure}
    \centering
    \includegraphics[width=0.5\textwidth]{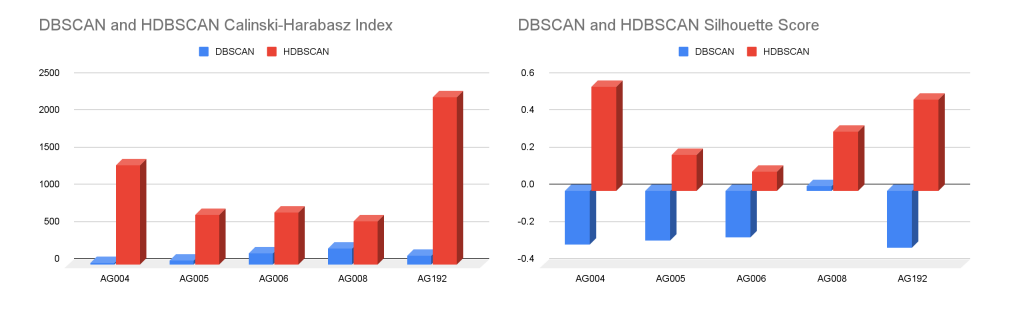}
    \caption{Performance Metrics For Clustering}
    \label{fig:3}
\end{figure}
It was observed that AGGLO, despite its capability to handle thin and elongated data shapes for clustering, did not present any advantages over other tested clustering methods. AGGLO was found to incorporate all noise present in the data, rendering DBSCAN and HDBSCAN more informative due to their resilience to noise (See Figure~\ref{fig:1}).

\begin{figure}[htbp]
    \centering
    \includegraphics[width=0.5\textwidth]{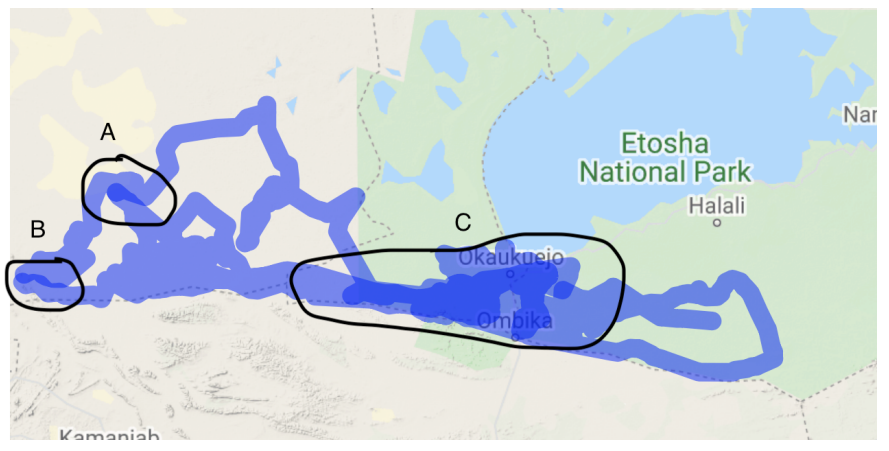}
    \caption{State Mapping on AG004}
    \label{fig:4}
\end{figure}
For the evaluation of clustering algorithms, two metrics were chosen: the Silhouette Score and the Calinski-Harabasz Index. These metrics were selected as they effectively gauge the density, separation, and variation of the clusters (See Figure~\ref{fig:3}). 
\begin{figure}
    \centering
    \includegraphics[height=5cm,width=0.5\textwidth]{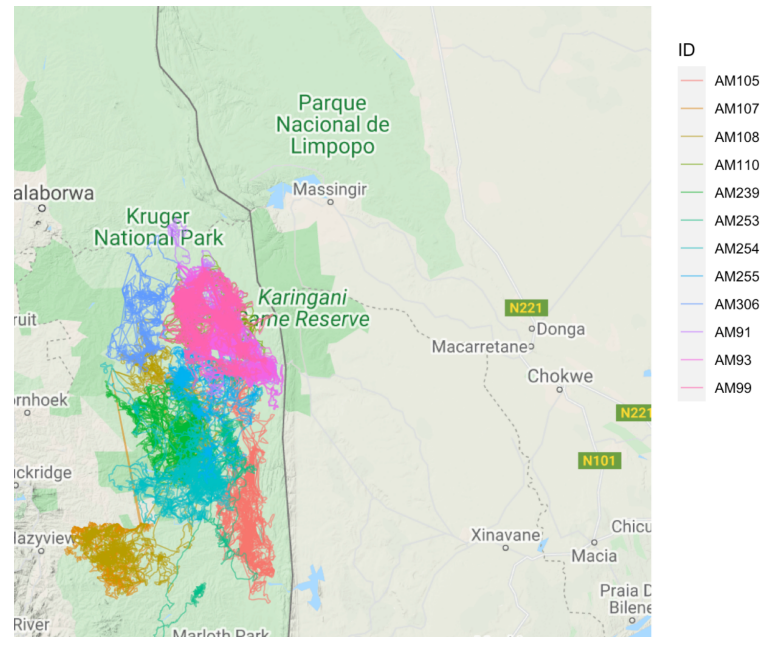}
    \caption{Kruger Elephant Groups Pathing Mapped}
    \label{fig:9}
\end{figure}
The Silhouette Score demonstrated significant differences in the performance of each clustering algorithm on each data set. 

The HDBSCAN algorithm did not return any negative coefficients across the clustering sets, and exhibited excellent scores for AG004 and AG192. 
It seems that the ability to detect varying density clusters in HDBSCAN offers a performance advantage. In terms of the Calinski-Harabasz index, HDBSCAN scored substantially higher across all elephant groupings, suggesting that HDBSCAN clusters were denser and more well-separated.
\begin{figure}
    \centering
    \includegraphics[width=0.5\textwidth]{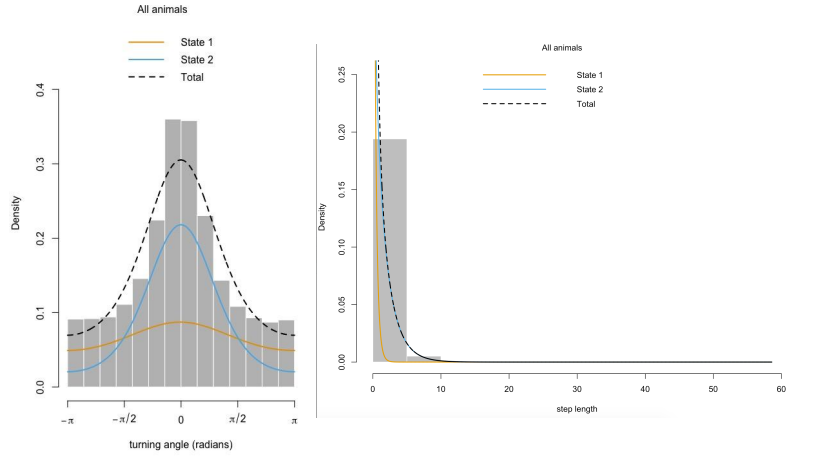}
    \caption{Elephant Movement Charecteristics}
    \label{fig:12}
\end{figure}
\begin{figure}
    \centering
    \includegraphics[width=0.5\textwidth]{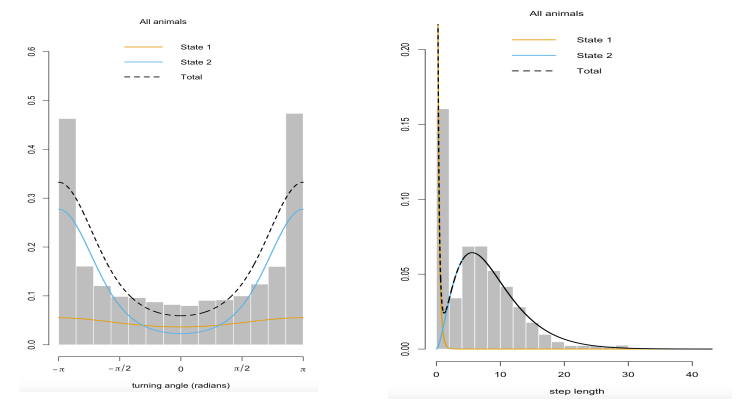}
    \caption{Jaguar Movement Charecteristics}
    \label{fig:13}
\end{figure}

\subsection{HMM Modeling}

Utilizing the MoveHMM R package, we performed Hidden Markov Model (HMM) analysis of elephant group movement. This model uses a bivariate time series of distance traveled between time (step size) and the change of direction between time (turning angle). 
\begin{figure}
    \centering
    \includegraphics[height=4cm,width=0.5\textwidth]{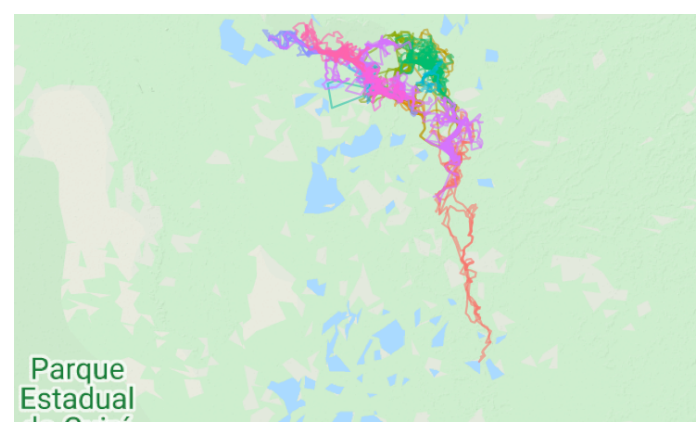}
    \caption{Brazil Jaguars Pathing}
    \label{fig:14}
\end{figure}
For the analysis, initial step and angle parameters were established for each state (See Figure~\ref{fig:4}). The plots generated for ``AG004" depict two states: a transit state characterized by larger steps with smaller turns, and a foraging state where the elephant takes shorter steps with larger turns.
\begin{figure}
    \centering
    \includegraphics[width=0.5\textwidth]{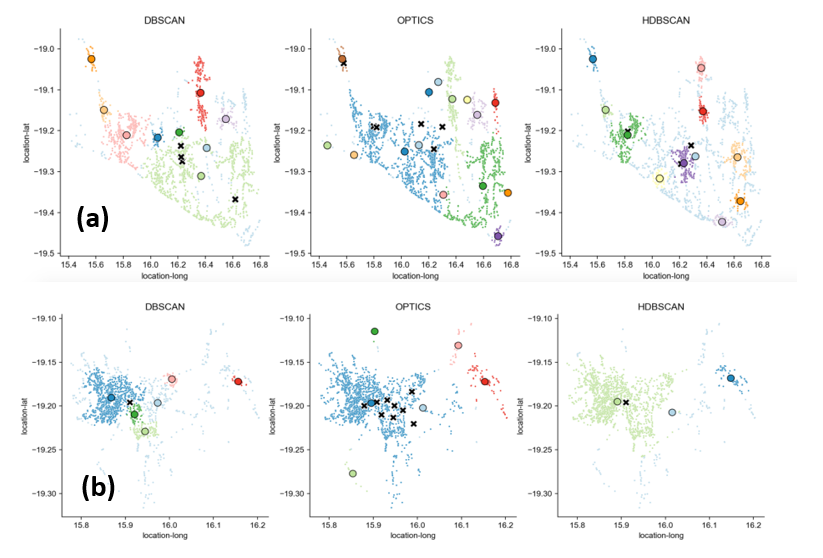}
    \caption{(a) Etosha National Park’s Wet Season as an average of all the years; (b) Etosha National Park’s Dry Season as an average of all the years}
    \label{tempclust}
\end{figure}
Mapping was conducted to produce state-wise mapping for all elephant groups at Etosha National Park (See Figure~\ref{fig:5}). These maps offer insights into key points of interest for all groupings of elephants and help identify locations of water and food resources. The transition probabilities of Etosha elephants (See Figure~\ref{fig:6}) and Kruger elephants (See Figure~\ref{fig:7}) were also examined. Figure~\ref{fig:9} depicts the movement path for elephant groups at Kruger National Park. The results indicated that changes in temperature play a smaller role in influencing the transition between states for Etosha elephants compared to Kruger elephants.
\par Analyzing two diagrams (See Figure~\ref{tempclust}(a) and Figure~\ref{tempclust}(b)) reveals a distinct contrast in elephant behavior between the wet and dry seasons. During the dry season, elephants exhibit a tendency to congregate and cluster together in specific locations, whereas in the wet season, they disperse and spread out over a wider area. This observation highlights a clear pattern of behavior based on the prevailing weather conditions. The dataset used for studying elephant movements in Etosha National Park encompasses a timeframe ranging from 2008 to 2014. Despite the span of years, a consistent pattern emerges, demonstrating that elephants have consistently exhibited the same behavior during every wet season and every dry season throughout the observed period. This consistent recurrence of behavior across multiple years underscores the reliability and repeatability of the observed pattern. By recognizing and understanding these seasonal behavioral patterns, we can gain valuable insights into elephant dynamics and potentially leverage this knowledge for various applications, such as conservation efforts, resource management, and human-elephant conflict mitigation strategies.
\subsection{Comparison with Jaguar Movement}
When comparing the movement patterns of different animal species, such as elephants and jaguars, it becomes evident that their behaviors are influenced by the characteristics of their respective environments. Here, we focus on scalability of animal movement and how it can shed light on habitats in which these animals reside.

In Kruger National Park, elephants exhibit a more linear form of movement, characterized by traveling between specific points. Figure~\ref{fig:12} illustrates this behavior, showing elephants with minimal turning angles and short step lengths. Such movement patterns can be attributed to the open safari landscape found in the park. As elephants are large herbivores, they spend a significant amount of their time traversing the terrain in search of resources. Their movement from one resource to another explains the observed linear paths they follow.

Contrasting the elephants' behavior, jaguars tracked on the largest floodplain in Brazil demonstrate distinct movement patterns. Figure~\ref{fig:13} displays the movement plots of jaguars, which exhibit random and erratic trajectories. Jaguars are known to exhibit high-degree turns and cover large distances in their movement, often adopting circular paths. The environment of the Brazilian floodplain plays a crucial role in shaping their movement behavior. Unlike elephants, the floodplain provides a consistent distribution of resources throughout, eliminating the need for jaguars to travel long distances or target specific areas for foraging. Figure~\ref{fig:14} further illustrates this point by showcasing the pathing of jaguars in Brazil.

Analyzing individual jaguars' movements reveals that they primarily occupy certain territorial ranges, with some exceptions. These ranges are evident from the mapping of each jaguar's path. However, there may be outliers, such as a particular jaguar that follows a distinct and precise path to and from a specific location. Additionally, jaguars tend to stay relatively close to bodies of water, indicating their dependence on these water sources. The presence of water serves as an anchor, restricting jaguars' movements to specific areas within their habitat.

It is worth noting that both elephant and jaguar movement patterns are influenced by the features of the land they inhabit. Elephants' extensive travel and pathing to specific areas are driven by the need to access resources, while jaguars' erratic movement and territorial ranges reflect the abundance of resources in their environment, supplemented by the presence of water sources. By studying and comparing the scalability of animal movement, we can gain insights into the unique characteristics of different ecosystems and the ways in which they shape animal behaviors. 
\subsubsection*{Consideration of a Control Species in Elephant Habitats:} The comparison between elephant and jaguar movements provides valuable insights about how species-specific behaviors and environmental characteristics influence animal mobility patterns. However, to isolate unique elephant behaviors, it may be beneficial to consider a `control' species cohabiting with elephants and having different propensities for human conflict. For example, less conflict-prone species such as giraffes or zebras could serve as controls. If these species demonstrate similar linear paths between resources as observed in elephants, it could suggest that these movement patterns are largely dictated by the shared environment and resource availability, rather than being exclusive to elephants. Contrasting movement patterns might indicate species-specific factors such as size or social structure influencing unique elephant behaviors. Alternatively, considering a species more prone to human conflict like lions or hyenas could provide interesting insights. Shared movement patterns highlight commonalities among species frequently encountering human conflict, while unique patterns in elephants could expose specific behaviors making them more susceptible to such conflicts. Such comparisons within the same habitat could help isolate species-specific movement patterns, providing better understanding of the unique factors influencing elephant movements.
\vspace{-0.5cm}
\subsection{Use-case with POI data} 
Multiple methods were employed to study elephant movement patterns. DBSCAN was utilized to calculate clusters and centroids based on location alone ("Without Temp-influenced") and in combination with temperature data ("Temp-influenced"). Both feature spaces demonstrated promising capabilities in identifying centroids in elephant movement, with each approach uncovering centroids that the other method missed. To augment datasets lacking temperature information, historical weather API was employed to incorporate temperature data, successfully enriching the data for centroid identification. Additionally, a "fuzzy" timestamp matching technique was applied to expand the availability of temperature data for analysis.

To ascertain the practical significance of these techniques in real-world applications, it is essential to overlay additional contextual information onto the calculated centroids. This includes factors such as terrain characteristics, satellite imagery, human settlements, water sources, and population density. By incorporating this supplementary data, we can assess whether the calculated centroids, both Temp-Influenced and Without Temp-influenced, hold any meaningful insights. To facilitate exploration of the calculated centroids and their relationship with these additional features, an application has been developed. The application allows users to examine the centroids interactively and visualize them in conjunction with the overlaid contextual information\footnote{The application can be accessed at \url{https://g1776-elephantcentroids-app-95v7am.streamlit.app/}}.
\begin{figure}
    \centering
    \includegraphics[width=0.5\textwidth]{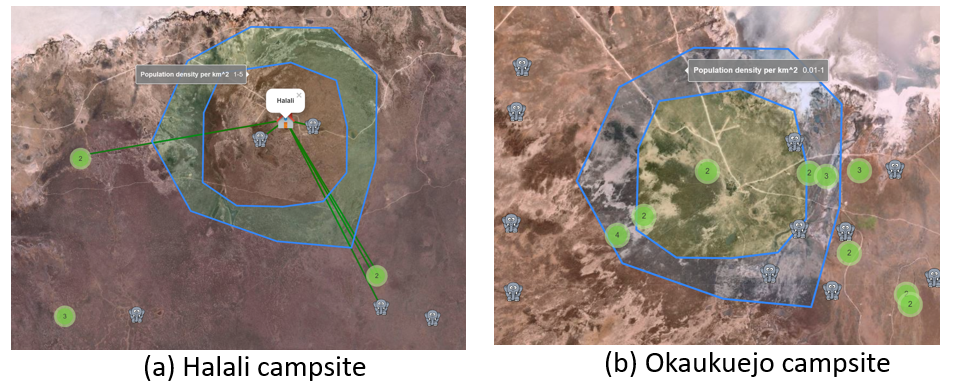}
    \caption{Elephant centroids around human settlement in Etosha National Park}
    \label{campsite}
\end{figure}
In Figure~\ref{campsite}, elephant centroids in the vicinity of human settlements within Etosha National Park, specifically at the Halali and Okaukuejo campsites, are depicted. The centroids are represented by elephant icons, with green markers indicating groups of centroids. These centroids encompass both Temp-Influenced and Without Temp-influenced variants, utilizing fuzzy matching. The circular boundary surrounding the settlement represents the extent of human population in the area. The presence of centroids within and around this boundary suggests the interest of elephants in the settlement and its surrounding region. According to information from Etosha National Park's website, both the Halali and Okaukuejo campsites feature watering holes, which could explain the elephants' affinity for these areas. Additionally, centroids are observed near bodies of water, where elephant centroids from the Kruger dataset align along a river, with a nearby campsite visible. This implies that elephants may prioritize the presence of rivers over human settlements, as the concentration of centroids is primarily along the river rather than around the settlement.

Analyzing the centroids within their surrounding context provided valuable insights into their potential applications. It was observed that some centroids were situated near settlements with watering holes, while others tended to be in proximity to rivers. Interestingly, no discernible pattern emerged to suggest that Temp-influenced centroids differed significantly from Without Temp-influenced centroids. We initially anticipated that Temp-influenced centroids would exhibit a greater affinity for water sources and clusters of trees (providing shade) compared to the Without Temp-influenced centroids. However, this expectation was not supported by the analysis. Overall, the contextual examination of the centroids allowed for a more comprehensive understanding of their practical utility. It highlighted the relationship between centroids and their surroundings, indicating that some centroids are found near settlements with watering holes, while others are concentrated along rivers. However, no distinctive pattern emerged to differentiate the Temp-influenced centroids from the Without Temp-influenced centroids in terms of their preferences for water sources or tree clusters.
\begin{figure}
    \centering
    \includegraphics[width=0.5\textwidth]{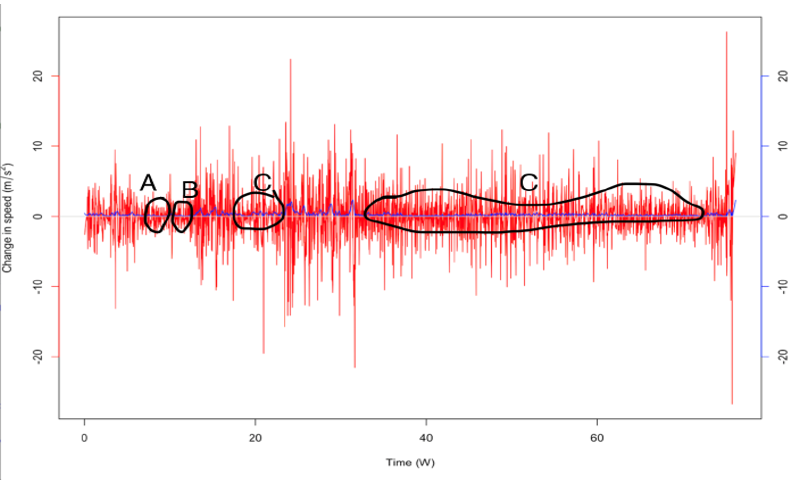}
    \caption{Velocity/ Acceleration of AG004}
    \label{speed}
\end{figure}
\begin{table}[htbp]
\centering
\renewcommand{\arraystretch}{1.2}
\resizebox{0.5\textwidth}{!}{
\begin{tabular}{lllr}
\toprule
\textbf{Geometry} & \textbf{Name} & \textbf{Type} & \textbf{\# Centroids in} \\ & & &  \textbf{settlement cluster} \\
\midrule
POINT (16.4710969 -19.0356338) & Halali & village & 23 \\
POINT (15.295068 -17.6750468) & Ogongo & village & 10 \\
POINT (14.5680119 -17.3940925) & Omahenene & village & 9 \\
POINT (16.0210914 -17.9842151) & Olukonda & hamlet & 9 \\
POINT (14.3015546 -19.8967052) & - & hamlet & 1 \\
POINT (15.9146617 -17.2211624) & Okawe & village & 1 \\
POINT (14.4043373 -18.7978984) & - & village & 1 \\
POINT (14.1598649 -18.6311834) & Otjitundua & village & 1 \\
POINT (14.960421 -20.37384) & Khorixas &  town & 1 \\
POINT (17.4722164 -17.6549121) & Omupini & hamlet & 1 \\
\bottomrule
\end{tabular}%
}
\caption{Settlement Clusters and Centroid Counts}
\label{tab:settlements}
\end{table}
\par The presence of black circles on the plot (See Figure~\ref{speed}) signifies that the elephants spend a significant amount of time in those specific areas. Among these circles, the one labeled as C is located near a man-made waterhole and a local tourist station. This finding suggests that the elephants are likely drawn to this location due to the availability of water and potentially other resources. The circled area labeled as B corresponds to the vicinity of Dolomite Camp, which serves as an explanation for its designation as an area of interest. It is plausible that the camp provides attractants or resources that make it appealing to the elephants. Regarding the circle labeled as A, its placement suggests that it may represent an area abundant in food, water, and/or shade. This inference is supported by the fact that the group of elephants traveled westward towards this particular location. It is likely that this area offers the necessary resources to support the elephants' needs during their journey. Overall, the generated plot effectively illustrates the areas of interest for the elephant group AG004. By analyzing the distribution of the black circles, we can identify key locations where the elephants spend a substantial amount of time. This information can be crucial for understanding their habitat preferences, resource utilization, and potentially informing conservation efforts and management strategies to ensure the long-term well-being of the elephant population.
\par By considering the combined centroids along with human settlements, we gain the capability to identify settlements that elephants may potentially inhabit at certain times. When visually inspecting the map, it becomes apparent which settlements have a significant concentration of elephant centroids, signaling the need for further investigation. To automate this selection process, we applied KMeans and DBSCAN algorithms to the problem. Subsequently, we ranked the settlements based on the number of elephant centroids present in each respective cluster. The outcome of this process, executed on the Etosha dataset, is presented in Table~\ref{tab:settlements}. This ranking provides a valuable starting point for directing efforts towards locating elephants.

Our research into elephant movement ecology suggests several prospective avenues for exploration, which may enhance conservation strategies. Understanding the impacts of climate change and land use changes on elephant migration, alongside the effects of extreme weather events, interactions with other species, landscape connectivity, behavioral alterations due to poaching, and responses to human infrastructure and changing rainfall patterns could present significant contributions. A vital focus could be the role of Artificial Intelligence (AI) and Machine Learning (ML) in analyzing extensive data from sources like GPS collaring, and predicting elephant movements based on varied factors like climate, land use, and poaching activities. AI and ML could also be instrumental in real-time monitoring and providing alerts about potential human-elephant conflicts or imminent threats like poaching, aiding quicker responses. These directions aim to encourage further research, underlining the significance of AI and ML tools in wildlife ecology and conservation, and fostering sustainable human-elephant coexistence in Sub-Saharan Africa.

\vspace{-0.3cm}
\section{Conclusion and Future Works}
This paper provides insightful findings on animal movement patterns, specifically elephants, demonstrating the influence of temperature and land features on their behavior. 
The methodologies used, when extended to analyze jaguar data, revealed distinct behaviors and movement strategies. 
In future, we plan to integrate uncertainty analysis into our predictive model as part of our future work. This includes developing methods to quantify and visualize error propagation within our predictions, especially considering geographical features and human settlement influences. 
Furthermore, we plan to incorporate measures to adjust the model's confidence bounds based on these error analyses, enabling the tool to self-calibrate as it encounters varying levels of uncertainty across different terrain types.

\section*{Acknowledgements}
This research was partially funded by the Federal Ministry of Education and Research (BMBF),
Germany under the project LeibnizKILabor with grant No. 01DD20003.
\bibliographystyle{ACM-Reference-Format}

\end{document}